\documentclass[10pt,conference]{./Components/MetaFiles/IEEEtran}

\usepackage{times,latexsym,psfrag,eufrak,dsfont}
\usepackage{epstopdf}
\usepackage{bbm}
\usepackage{siunitx}

\usepackage{gensymb}
\usepackage{amsmath,amssymb,amsfonts,amsthm}
\usepackage{epsfig}
\usepackage{cite}
\usepackage{hhline}
\usepackage{multirow}
\usepackage{xcolor}
\usepackage{makecell}
\usepackage{subcaption}
\usepackage{capt-of}
\usepackage[labelformat=simple]{subcaption}

\usepackage{enumerate}
\usepackage{enumitem}
\usepackage{color}
\usepackage{graphicx}
\usepackage{algorithmic}
\usepackage[ruled]{algorithm}
\usepackage{booktabs}
\usepackage{bm}
\usepackage[nolist,printonlyused]{acronym} 
\usepackage{colortbl}
\usepackage{tabularx}
\usepackage{float}
\newcolumntype{Y}{>{\centering\arraybackslash}X}

\usepackage[T1]{fontenc}
\usepackage[utf8]{inputenc}
\usepackage{comment} 
\usepackage[author={Hossein}]{pdfcomment}

\usepackage{pgf}
\usepackage{pgfplots}
\usepackage{tikz}
\usepackage{grffile}
\pgfplotsset{compat=newest}
\usetikzlibrary{plotmarks}
\usepgfplotslibrary{patchplots}
\usetikzlibrary{automata,arrows,positioning,calc,matrix,decorations.markings,decorations.pathreplacing,patterns,arrows.meta}
 \pgfdeclarepatternformonly{south west lines}{\pgfqpoint{-0pt}{-0pt}}{\pgfqpoint{3pt}{3pt}}{\pgfqpoint{3pt}{3pt}}{
        \pgfsetlinewidth{0.4pt}
        \pgfpathmoveto{\pgfqpoint{0pt}{0pt}}
        \pgfpathlineto{\pgfqpoint{3pt}{3pt}}
        \pgfpathmoveto{\pgfqpoint{2.8pt}{-.2pt}}
        \pgfpathlineto{\pgfqpoint{3.2pt}{.2pt}}
        \pgfpathmoveto{\pgfqpoint{-.2pt}{2.8pt}}
        \pgfpathlineto{\pgfqpoint{.2pt}{3.2pt}}
        \pgfusepath{stroke}}
\makeatletter
\newcommand{\vast}{\bBigg@{3}}
\newcommand{\Vast}{\bBigg@{4}}
\makeatother
\renewcommand{\IEEEQED}{\IEEEQEDopen}

\showoutput
\IEEEoverridecommandlockouts

\newtheorem{prop}{Proposition}
\newtheorem{lem}{Lemma}

\newtheorem{defi}{Definition}
\newtheorem{rmk}{Remark}

\def\({\left(}
\def\){\right)}

\def\[{\left[}
\def\]{\right]}

\newcommand{\be}{\begin{equation}}
\newcommand{\ee}{\end{equation}}
\newcommand{\ba}{\begin{array}}
\newcommand{\ea}{\end{array}}
\newcommand{\bea}{\begin{eqnarray}}
\newcommand{\eea}{\end{eqnarray}}

\newcommand{\vbar}{\raisebox{.17ex}{\rule{.04em}{1.35ex}}}
\newcommand{\vbarind}{\raisebox{.01ex}{\rule{.04em}{1.1ex}}}
\newcommand{\D}{\ifmmode {\rm I}\hspace{-.2em}{\rm D} \else ${\rm I}\hspace{-.2em}{\rm D}$ \fi}
\newcommand{\T}{\ifmmode {\rm I}\hspace{-.2em}{\rm T} \else ${\rm I}\hspace{-.2em}{\rm T}$ \fi}
\newcommand{\B}{\ifmmode {\rm I}\hspace{-.2em}{\rm B} \else \mbox{${\rm I}\hspace{-.2em}{\rm B}$} \fi}
\newcommand{\Hil}{\ifmmode {\rm I}\hspace{-.2em}{\rm H} \else \mbox{${\rm I}\hspace{-.2em}{\rm H}$} \fi}
\newcommand{\Cind}{\ifmmode \hspace{.2em}\vbarind\hspace{-.25em}{\rm C} \else \mbox{$\hspace{.2em}\vbarind\hspace{-.25em}{\rm C}$} \fi}
\newcommand{\Q}{\ifmmode \hspace{.2em}\vbar\hspace{-.31em}{\rm Q} \else \mbox{$\hspace{.2em}\vbar\hspace{-.31em}{\rm Q}$} \fi}
\newcommand{\Z}{\ifmmode {\rm Z}\hspace{-.28em}{\rm Z} \else ${\rm Z}\hspace{-.38em}{\rm Z}$ \fi}

\makeatletter
\let\save@mathaccent\mathaccent
\newcommand*\if@single[3]{%
  \setbox0\hbox{${\mathaccent"0362{#1}}^H$}%
  \setbox2\hbox{${\mathaccent"0362{\kern0pt#1}}^H$}%
  \ifdim\ht0=\ht2 #3\else #2\fi
  }
\newcommand*\rel@kern[1]{\kern#1\dimexpr\macc@kerna}
\newcommand*\widebar[1]{\@ifnextchar^{{\wide@bar{#1}{0}}}{\wide@bar{#1}{1}}}
\newcommand*\wide@bar[2]{\if@single{#1}{\wide@bar@{#1}{#2}{1}}{\wide@bar@{#1}{#2}{2}}}
\newcommand*\wide@bar@[3]{%
  \begingroup
  \def\mathaccent##1##2{%
    \let\mathaccent\save@mathaccent
    \if#32 \let\macc@nucleus\first@char \fi
    \setbox\z@\hbox{$\macc@style{\macc@nucleus}_{}$}%
    \setbox\tw@\hbox{$\macc@style{\macc@nucleus}{}_{}$}%
    \dimen@\wd\tw@
    \advance\dimen@-\wd\z@
    \divide\dimen@ 3
    \@tempdima\wd\tw@
    \advance\@tempdima-\scriptspace
    \divide\@tempdima 10
    \advance\dimen@-\@tempdima
    \ifdim\dimen@>\z@ \dimen@0pt\fi
    \rel@kern{0.6}\kern-\dimen@
    \if#31
      \overline{\rel@kern{-0.6}\kern\dimen@\macc@nucleus\rel@kern{0.4}\kern\dimen@}%
      \advance\dimen@0.4\dimexpr\macc@kerna
      \let\final@kern#2%
      \ifdim\dimen@<\z@ \let\final@kern1\fi
      \if\final@kern1 \kern-\dimen@\fi
    \else
      \overline{\rel@kern{-0.6}\kern\dimen@#1}%
    \fi
  }%
  \macc@depth\@ne
  \let\math@bgroup\@empty \let\math@egroup\macc@set@skewchar
  \mathsurround\z@ \frozen@everymath{\mathgroup\macc@group\relax}%
  \macc@set@skewchar\relax
  \let\mathaccentV\macc@nested@a
  \if#31
    \macc@nested@a\relax111{#1}%
  \else
    \def\gobble@till@marker##1\endmarker{}%
    \futurelet\first@char\gobble@till@marker#1\endmarker
    \ifcat\noexpand\first@char A\else
      \def\first@char{}%
    \fi
    \macc@nested@a\relax111{\first@char}%
  \fi
  \endgroup
}
\makeatother

\def\ba{{\bf a}}

\def\be{{\bf e}}

\def\tt{{\mathrm{t}}}

\def\Nb{N_{\text{BS}}}
\def\Nu{N_{\text{UE}}}
\def\Thetab{\theta_{\text{BS}}}
\def\Thetau{\theta_{\text{UE}}}
\def\Gb{G_{\text{BS}}}
\def\Gu{G_{\text{UE}}}

\def\b1{{\bf 1}}

\renewcommand{\IEEEQED}{\IEEEQEDopen}

\newcommand{\etal}[1]{{#1}{~\textit{et al.}}}

\begin{document}
\graphicspath{{Components/Figures/}}

\title{Reducing Initial Cell-search Latency in \\  mmWave Networks}

\author{\IEEEauthorblockN{Yanpeng Yang, Hossein S. Ghadikolaei, Carlo Fischione, Marina Petrova, and Ki Won Sung}
\IEEEauthorblockA{School of Electrical Engineering and Computer Science, KTH Royal Institute of Technology, Stockholm, Sweden\\
E-mails: \{yanpeng, hshokri, carlofi, petrovam, sungkw\}@kth.se}
}

\maketitle

\begin{abstract}
Millimeter-wave (mmWave) networks rely on directional transmissions, in both control plane and data plane, to overcome severe path-loss. Nevertheless, the use of narrow beams complicates the initial cell-search procedure where we lack sufficient information for beamforming. In this paper, we investigate the feasibility of random beamforming for cell-search. We develop a stochastic geometry framework to analyze the performance in terms of failure probability and expected latency of cell-search. Meanwhile, we compare our results with the naive, but heavily used, exhaustive search scheme. Numerical results show that, for a given discovery failure probability, random beamforming can substantially reduce the latency of exhaustive search, especially in dense networks. Our work demonstrates that developing complex cell-discovery algorithms may be unnecessary in dense mmWave networks and thus shed new lights on mmWave system design.
\end{abstract}


\section{Introduction}
Millimeter-wave (mmWave) technology is one of the essential components of future wireless networks to support extremely high data rate services~\cite{Rappaport2013}. To compensate for the severe path-loss, mmWave systems rely on directional transmissions using large antenna arrays both at the transmitter and at the receiver. Such directional transmission, albeit reduces the interference footprint and simplifies the scheduling task~\cite{Shokri2015Transitional}, complicates the initial synchronization and cell-search procedure as the antenna patterns of base stations (BS) --or access points-- and user equipment (UE) should be aligned to be able to receive time-frequency synchronization signals.

\etal{Barati}~\cite{Barati2016} proposed a link-level procedure for directional initial access in a single cell scenario. The synchronization signals and the detection approach is such that time-frequency-spatial synchronization occurs jointly. In contrast, \cite{Shokri-Ghadikolaei2015} proposed a two-level synchronization architecture for a multi-cell scenario where macro-level signals help BSs and UEs synchronize in time and frequency domains, and a local beam scanning executes the synchronization in the spatial domain. \cite{Li2017a} and \cite{Li2017} provide a system-level analysis of this scheme for a spacial beam scanning approach: exhaustive search. In particular, both BSs and UEs sequentially scan the whole angular space to find ``the best'' link quality. However, finding ``an acceptable'' link, formally defined in Section~\ref{sec: system-model}, may be much faster, reducing the latency for the establishment of data plane. Fast initial access becomes even more important for wireless sensor networks comprising many wake-up radios in which the synchronization may take longer than the actual data transmission, leading to a poor latency performance. As we show throughout this paper, increasing the BS (or access point) density can reduce the initial access latency, as opposed to the exhaustive-search.

Random beamforming is an alternative to the exhaustive search in which the BSs (and UEs) focus their antenna patterns toward a randomly pick direction.
\etal{Lee}~\cite{Lee2016a} investigated the data plane performance (in terms of sum-rate) of random beamforming. References\cite{Shokri-Ghadikolaei2015} and \cite{Abu-Shaban2016} have shown the feasibility of applying random beamforming for initial cell search of mmWave networks. In particular, \cite{Shokri-Ghadikolaei2015} analyzed the delay statistics of initial access, assuming a noise-limited scenario and a deterministic channel model, and showed promising results for the performance of random beamforming. \cite{Abu-Shaban2016} achieved a similar conclusion by analyzing the Cram\'{e}r-Rao lower bound for estimating directions of arrival and departure (essentially spatial synchronization). The analysis, however, is limited to a single-link scenario.


Motivated by these initial results, we substantially extend \cite{Shokri-Ghadikolaei2015} and \cite{Abu-Shaban2016} and provide a system-level framework to analyze the performance of initial access based on random beamforming in a multi-cell mmWave network. The main contributions of this paper are:
\begin{itemize}
\item By a stochastic geometry analysis, we evaluate the detection failure probability and delay distribution of random beamforming for initial cell search.
\item We characterize the tradeoff between failure probability and expected latency and show the superior latency performance of the random beamforming compared to the exhaustive search. This performance gain gets more prominent in dense mmWave networks.
\item We formulate an optimization problem to find the optimal beamwidth subject to a minimum detection failure probability.
\item We numerically analyze the performance of random beamforming for cell-search and demonstrate that it may be unnecessary to develop a complicated cell-search algorithm in dense mmWave networks.
\end{itemize}

The rest of the paper is organized as follows. In Section~\ref{sec: system-model}, we describe the system model. Main results are presented in Section~\ref{sec: main-results}, followed by numerical performance evaluation in Section~\ref{sec: Numerical-results}. The paper is concluded in Section~\ref{sec: conclusions}.

\begin{figure}[t]
\centering
\includegraphics[width=0.95\columnwidth]{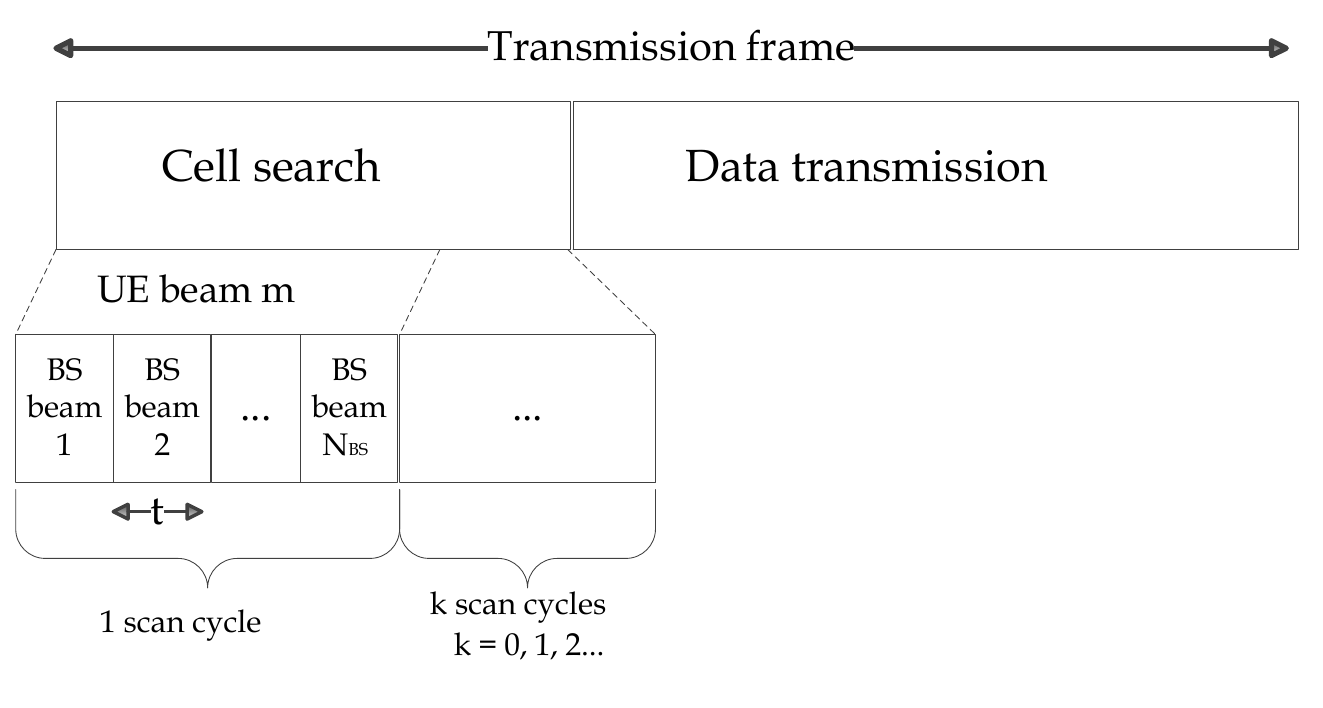}

\caption{Illustration of transmission frame with random beamforming.}
\label{fig:illu}
\end{figure}

\section{System Model}\label{sec: system-model}
\subsection{Network and Antenna Models}
We consider a mmWave cellular network where the BSs are distributed according to a two-dimensional Poisson point processes (PPP) $\Phi \triangleq \{x_i\}$ with density $\lambda$. The UEs follow another independent PPP, from which the typical UE, located at the origin, is our focus according to the Slivnyak's theorem~\cite{Haenggi2012}.


Each BS and UE is equipped with multiple antennas and supports analog beamforming. We believe that digital or hybrid beamforming does not suit initial cell-search due to the existence of many antenna elements and lack of prior channel knowledge, translated into the need for costly pilot transmission schemes. We consider half-power beamwidths of $\Thetab$ and $\Thetau$ at the BSs and UEs, respectively, with the corresponding antenna gains $\Gb$ and $\Gu$. For analytical tractability, we model the actual antenna patterns by a sectorized antenna model as in~\cite{Shokri-Ghadikolaei2015}. In an ideal sectorized antenna pattern, the antenna gain $G_{\mathrm x}$, ${\mathrm x} \in \{{\text{BS}}, {\text{UE}}\}$, as a function of beamwidth $\theta_{\mathrm x}$ is a constant in the main lobe and a smaller constant in the side lobe, given by
\begin{align}\label{equ:antenna}
G_{\mathrm x}(\theta_{\mathrm x})=\left\{
        \begin{array}{ll}
          \frac{2\pi-(2\pi-\theta_{\mathrm x})\epsilon}{\theta_{\mathrm x}}, & \text{in the main lobe,} \\
          \epsilon, & \text{in the side lobe,} \\
        \end{array}
      \right.
\end{align}
where typically $\epsilon \ll 1$. For sake of mathematical simplicity, we assume $\epsilon=0$, i.e., no sidelobe gain, and only one RF chain at the BSs and UEs, though the analysis can be readily extended for the general case.

For a given $\Thetab$ and $\Thetau$, which are a non-increasing function of the number of antenna elements, BSs and UEs sweep the entire angular space by $\Nb = \lceil2\pi/\Thetab\rceil$ and $\Nu = \lceil2\pi/\Thetau\rceil$ beamforming vectors, respectively. Without loss of generality of the main conclusions, we assume that $2\pi/\Thetab$ and $2\pi/\Thetab$ are integers and drop $\lceil\cdot\rceil$ operator.

\subsection{Propagation and Blockage Models}
We adopt a simple distance-dependent attenuation model with path-loss exponent $\alpha > 2$ as \cite{Shokri-Ghadikolaei2015} and \cite{Li2017}. The path-loss from a BS located at $x_i$ to the typical UE, located at the origin, is
\begin{equation*}
\ell(\lVert x_i \rVert)=\(\frac{c}{4 \pi \lVert x_i \rVert f_c}\)^{\alpha} \:,
\end{equation*}
where $c$ is the light speed, $f_c$ is the operating frequency and $\Vert x_i \Vert$ represents the distance between $x_i$ and the origin.

Let $S_{i}$ be a binary variable taking 1 iff there is a LoS condition between BS $x_i$ and the typical UE. We apply an exponential blockage model~\cite{bai}. Formally, $S_{i}=1$ with probability $e^{-\beta \lVert x_i \rVert}$, where $\beta$ is a constant parameter determined by the size and density of the obstacles, and $\beta^{-1}$ represents the average length of a LoS link. Similar to \cite{Park2016}, we assume that mmWave signals cannot penetrate blockages. This assumption is also motivated by \cite{Shokri2015IMSindex} which shows that neglecting NLoS links comes at almost no penalty in the accuracy of statistical analysis. The impact of NLoS links is left for a future work.

Small-scale fading $h_i$ between BS $x_i$ and the typical UE follow a unit-mean Rayleigh distribution. Compared to more realistic models for LoS paths such as Nakagami fading, Rayleigh fading provides very similar design insights while leads to more tractable results~\cite{Li2017a}.

Denoting $p_{\text{BS}}$ as the BS transmit power and $W$ as the thermal noise, the signal-to-interference-plus-noise ratio (SINR) when the typical UE is receiving from BS $x_i$ is given by
\begin{equation*}
\text{SINR}_i =\frac{h_{i} \ell(\lVert x_i \rVert) S_{i}}{\sum\limits_{x_j \in \Phi \backslash x_i} h_{j} \ell(\lVert x_j \rVert)  S_{j} + \sigma^2} \:,
\end{equation*}
where $\sigma^2=W(p_{\text{BS}} \Gb \Gu)^{-1}$ is the normalized noise power.

\subsection{Random Beamforming Model}
The transmission frame under random beamforming is illustrated in Fig.~\ref{fig:illu}. The cell-search period comprises of several mini-slots. In each mini-slot, every BS independently and uniformly at random picks a direction out of $\Nb$. We define a scan cycle as the period within which every BS sends cell-search pilots to $\Nb$ directions. In each scan cycle, the UE antenna points to a random direction out of $\Nu$, and the BS covers all non-overlapping $\Nb$ directions. Different from exhaustive search in which the BS and UE need to cover all the $\Nb \Nu$ possible directions \cite{Li2017a}, the cell-search period of random beamforming can be dynamically adjusted. Once the UE received a pilot signal that meets a predefined SINR threshold, it is associated to the corresponding BS. Note that it may not be the final association of that UE, but once the UE is registered to the network, it can establish data plane, and the reassociation phase (to the best BS) could be smooth without service interruption~\cite{Shokri-Ghadikolaei2015}.

Let $t$ denote the duration of one mini-slot. Unlike \cite{Li2017a} where a fixed $t$ is considered, our mini-slot duration depends on the beamwidths and therefore on $\Nb$ and $\Nu$. This is a more realistic model because communication with narrower beams (higher antenna gains) requires less time than that with wider beams to ensure collecting the same energy at the receiver. Nevertheless, there exist a minimum value of mini-slot duration $t_0$ which corresponds to the minimum length of the symbols (e.g., the duration of one OFDM symbol). Using the simulation parameters of~\cite{Giordani2016}, we set $\Nb=12$ ($\Thetab = \ang{30}$) and $\Nu=4$ ($\Thetau = \ang{90}$) as our baseline to achieve $t_0$ mini-slot duration. Namely, $t=t_0$ when $\Nb \Nu = 48$, and in general it is inversely proportional to $\Nb \Nu$. For a general $\Nb$ and $\Nu$, the mini-slot duration is
\begin{equation}\label{eq: t}
t=\max(t_0,\frac{48 t_0}{\Nb\Nu}) \:,
\end{equation}
and the number of search mini-slots in every scan cycle is $\Nb$.

\subsection{Performance Metrics}
We say the typical UE successfully detects the cell if the strongest signal it receives in any mini-slot from one of the directions achieves a minimum SINR threshold $T$. Namely, the success event is $\mathbb{I}\{\max_{x_i \in \Phi} \text{SINR}_{i} \geq T \}$, where $\mathbb{I}\{\cdot\}$ is the indicator function. For any realization of the topology $\Phi$ and channel fading $h$, detection failure may happen due to two reasons: there is no BS inside the UE's main beam or the ones inside cannot meet the detection threshold $T$.

Now, we define two performance metrics, for the typical UE, evaluated throughout the paper.
\begin{defi}
The detection failure probability $P_f(N_c)$ is the probability that the UE is not detected by any BS within $N_c$ mini-slots. \end{defi}

\begin{defi}
Given a time-budget of $N_c$ mini-slots, the cell-search latency $L(N_c)$ of a UE is defined as the time period by which the UE successfully detects a cell-search pilot and can be registered with the corresponding BS.
\end{defi}
\begin{rmk}
For all $N_c \in \mathds{N}$, we have $t_0 \leq L(N_c) \leq N_c t$.
\end{rmk}
The lower-bound of $L(N_c)$ is due to the need for sending at least one pilot. Note that smaller $N_c$ values correspond to smaller $L(N_c)$, and that $L(1) = t$, but the price is higher failure probability. In fact, reducing $N_c$ means limiting the search budget and registering only a few users that can detect pilot signals within $N_c$ mini-slots. $L(\infty)$ characterizes the latency of UEs that can be detected ultimately. In some realizations of the random network topology, there is no BS close enough to the origin to detect the typical UE with the target SINR threshold. In those cases, the latency is undefined as the UE is fundamentally unable to find the pilot signals, independent of the cell-search policy. We formally characterize this issue in Lemma~\ref{lem: pfasymptotic}.

\section{Main results}\label{sec: main-results}
In this section, we present our main results and some insights on the performance of initial access using random beamforming. Proofs are provided in Appendix.

\begin{lem}\label{lem: pfasymptotic}
The probability that the typical UE has no LoS link to any BS is
\begin{align}\label{eq: Pnolos}
P_{\text{no-LoS}} = \exp(-\frac{2\lambda \pi}{\beta^2}) \:.
\end{align}
\end{lem}

\begin{prop}\label{prop: detection-failure-prob}
The detection failure probability $P_f(N_c)$ of a UE is
\begin{equation}\label{eq: Pf}
P_f= \max \((1- P_s)^{N_c}, P_{\text{no-LoS}}\) \:,
\end{equation}
where $P_{\text{no-LoS}}$ is given in \eqref{eq: Pnolos}, and $P_s$ is the successful detection probability in one mini-slot, given by
\begin{multline*}
P_s = \int_{0}^{\infty} \hspace{-2mm} e^{-T r^{\alpha} \sigma^2} \exp\(-\frac{2\pi}{\Nb \Nu} \lambda_b \int_0^{\infty} \frac{T r^{\alpha}e^{-\beta v}}{v^{\alpha}+T r^{\alpha}} v\mathrm{d}v\) \\ e^{-\beta r} \frac{2\pi}{\Nb \Nu} \lambda r \mathrm{d}r \:.
\end{multline*}
\end{prop}

Next, we derive the expected search latency, normalized to the mini-slot duration $t$. Proposition~\ref{prop: expected-latency} shows the average number of mini-slots required for detection a cell-search signal.
\begin{prop}\label{prop: expected-latency}
The normalized expected search latency given that the UE can be detected over $N_c$ mini-slots is
\begin{align}\label{eq: expected-latency}
\frac{\mathbb{E}[L(N_c)]}{t} = \frac{1-(N_c+1)(1-P_s)^{N_c}+N_c (1-P_s)^{N_c + 1}}{(1-P_f)P_s} \:.
\end{align}
\end{prop}

\begin{rmk}\label{rmk: asymptotic-latency}
When $N_c \to \infty$, the search latency $L(N_c)$ becomes geometrically distributed, implying that $\lim_{N_c \to \infty} t^{-1}\mathbb{E}[L(N_c)] \to P_s^{-1}$.
\end{rmk}

Proposition~\ref{prop: detection-failure-prob} and \ref{prop: expected-latency} characterize two important aspects of the cell-search under random beamforming. Increasing the search budget $N_c$ reduces the failure probability while increases the average search-latency. In the following, we optimize this tradeoff.

Consider the characterizations of the detection failure probability and latency, formulated in~\eqref{eq: Pnolos}, \eqref{eq: Pf}, and \eqref{eq: expected-latency}. We aim to design the BS beamwidth ($\Thetab$ or equivalently $\Nb$) to minimize the cell search latency given a failure probability constraint $P_f^{\text{max}} \in [0,1]$. Recall that $\Thetab= 2\pi/\Nb$. Moreover, from Fig.~\ref{fig:illu}, each scan cycle has $\Nb$ mini-slots. The optimal number of sectors for $k\in\mathbb{N}$ scan cycles (so $N_c = k\Nb$) is
\begin{subequations}\label{eq: optimization-problem}
\begin{alignat}{3}
\min_{\Nb}&  \hspace{3mm} && \mathbb{E}[L(k \Nb)]  \\
\text{s.t.} & && P_f(k \Nb) \leq P_f^{\text{max}}, \\
& && \Nb \in \mathbb{N} \:,
\end{alignat}
\end{subequations}
where $P_f^{\text{max}} $ and $k$ are inputs to this optimization problem. The problem can be utilized for system design in mmWave networks. With the knowledge of network deployment (BS desntiy), we can set the configuration of BS antennas to meet the requirements of various applications with different reliability and/or latency constraints. An example of the solutions to the problem can be observed from the figures in the next section. We finish this section by characterizing the performance of the exhaustive search.

\begin{rmk}\label{rmk: exhastive-search}
For the exhaustive search, given any $\Nb$ and $\Nu$, the search space of one scan cycle is $N_c = \Nb \Nu$ mini-slots. The detection failure probability is $P_f (N_c) = \max \((1- P_s)^{\Nb \Nu} , P_{\text{no-LoS}}\)$ and the expected search latency is $ \mathbb{E}[L (N_c)]=\Nb \Nu t$.
\end{rmk}

To illustrate Remark~\ref{rmk: exhastive-search}, we note that the exhaustive search has a fixed duration for scanning cycles ($\Nb \Nu$ mini-slots). If the UE is not detected in one cycle, which may happen due to random deep fading or random strong interference, it might be detected in the next cycles (in next frame perhaps). However, the UE cannot terminate the search process right after detecting one cell-search signal, as it is looking for the ``best'' one. In random beamforming, however, the UE is looking for a ``sufficiently good'' one and will probably terminate the process sooner than $\Nb \Nu$ mini-slots. In the next section, we show that this stop policy substantially reduces the cell-search latency.

\section{Numerical Results}\label{sec: Numerical-results}
In this section, we present the numerical results of cell search under random beamforming. The simulation parameters are summarized in Table~\ref{tab:para}. Unless otherwise specified, we set $\lambda = 10^{-3}/\text{m}^2$, $\Thetab = \ang{30}$ ($\Nb=12$), $\Thetau= \ang{90}$ ($\Nu=4$). As we mentioned in Introduction, state-of-the-art cell-search algorithms either work on the link-level or are designed for a single-cell scenario, which are not fairly comparable with our system-level multi-cell scenario.
The closest ones to our work are \cite{Li2017a} and \cite{Li2017}, which investigate exhaustive search with mini-slots of constant duration. To compare our proposed random beamforming with these works from the literature, we consider an exhaustive search that admits mini-slot duration based on \eqref{eq: t}.

\begin{table}
\caption{Main simulation parameters.}
\centering
\begin{tabular}{>{\centering}m{2cm} >{\centering}m{3cm} m{2cm}<{\centering}}
\hline \hline
Parameter & Description & Value \\
\hline
$p_{\text{BS}}$ & BS transmit power & 30 dBm\\
$\alpha$ & Path-loss exponent  & 3\\
$\beta$  & Blockage exponent   & 0.02\\
T & SINR threshold & 0 dB\\
$f_c$ & Operating frequency  & 28GHz \\
B & Control plane bandwidth & 1MHz\\
NF & Noise figure & 7dB\\
W & Thermal noise & -174 dBm/Hz \\\hline \hline
\end{tabular}
\label{tab:para}
\end{table}

\begin{figure}[t]
\centering
\begin{subfigure}[a]{0.5\textwidth}
 \centering
\footnotesize{\input{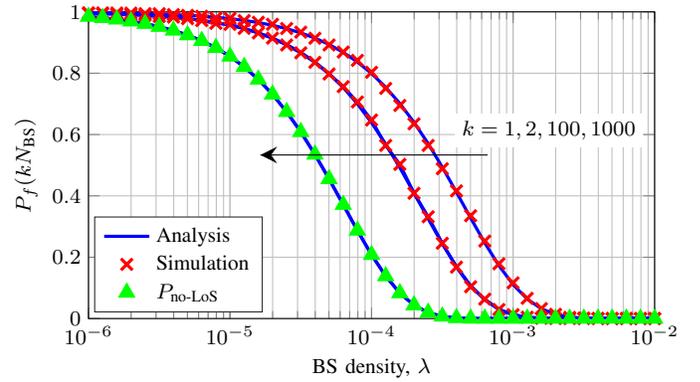}}
\caption{Detection failure probability.}
\label{fig:Pfdensity}
\end{subfigure}
\begin{subfigure}[a]{0.5\textwidth}
\centering
\footnotesize{\input{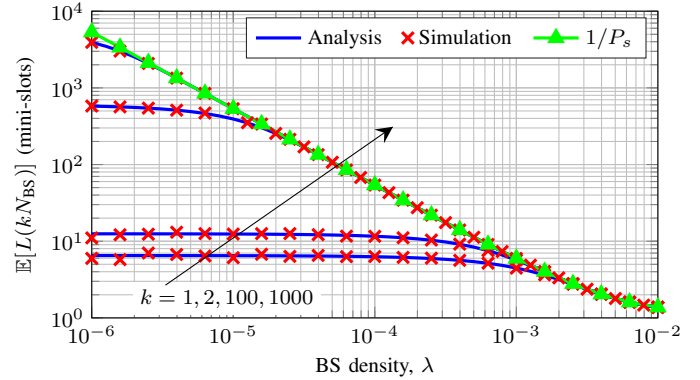}}
\caption{Expected searching latency.}
\label{fig:Nddensity}
\end{subfigure}
\caption{Effect of BS density $\lambda$ on the performance metrics ($\Nb = 12$). Detection failure probably for $k=100$ and $k=1000$ coincide with $P_{\text{no-LoS}}$.}
\label{fig:Density}
\end{figure}

Fig.~\ref{fig:Density} shows the cell-search performance against BS density $\lambda$. We observe a good matching between mathematical analysis and Monte Carlo simulations. From Fig.~\ref{fig:Pfdensity}, the detection failure probability reduces with the BS density. Besides, $P_f$ converge to $P_{\text{no-LoS}}$ as $k\to \infty$, shown in Proposition~\ref{prop: detection-failure-prob}. From Fig.~\ref{fig:Nddensity}, the expected search latency decreases for denser networks. The main reason is the availability of more candidate BSs to register the UE. Consistent with Remark~\ref{rmk: asymptotic-latency}, the latency converge to $P_{s}^{-1}$ as $N_c \to \infty$. Noting that the best possible values for the failure probability and normalized expected latency are 0 and 1, respectively. By employing random beamforming, we can get close to those values in dense mmWave networks.

\begin{figure}
\footnotesize{\begin{tikzpicture}
\begin{axis}[%
width=0.83\columnwidth,
height=0.45\columnwidth,
at={(0\columnwidth,0\columnwidth)},
scale only axis,
xmin=1,
xmax=20,
xtick={0,4,8,12,16,20},
xlabel={Number of sectors at BS, $\Nb$},
ymin=0,
ymax=80,
ylabel={$\mathbb{E}[L]/t_0$},
ylabel near ticks,
xmajorgrids,
ymajorgrids,
axis background/.style={fill=white},
legend style={at={(0.01,0.98)},anchor=north west,legend cell align=left, align=left,draw=white!15!black}
]
\addplot [color=blue, line width=1.1pt, mark size=2.2pt, mark=*, mark options={solid, fill=blue, blue}]
  table[row sep=crcr]{%
1	48\\
2	48\\
3	48\\
4	48\\
5	48\\
6	48\\
7	48\\
8	48\\
9	48\\
10	48\\
11	48\\
12	48\\
13	52\\
14	56\\
15	60\\
16	64\\
17	68\\
18	72\\
19	76\\
20	80\\
};
\addlegendentry{Exhaustive search}

\addplot [smooth, color=red, line width=1.1pt, mark size=2.6pt, mark=triangle*, mark options={solid, fill=red, red}]
  table[row sep=crcr]{%
1	21.1633118973712\\
2	14.1174736198738\\
3	11.2411246020549\\
4	9.65210069530581\\
5	8.63462549396533\\
6	7.92221956432752\\
7	7.39255518196729\\
8	6.98143316183794\\
9	6.65183588118301\\
10	6.38086905491029\\
11	6.15358293189132\\
12	5.95978289368704\\
13	6.27495518575348\\
14	6.58675763378838\\
15	6.89556363036963\\
16	7.20168355216436\\
17	7.50537887176606\\
18	7.80687239650603\\
19	8.10635586971599\\
20	8.40399573595256\\
};
\addlegendentry{Random beamforming}
\end{axis}
\end{tikzpicture}%
}

\caption{Expected latency of random beamforming and exhaustive search.}
\label{fig:delaycomp}
\end{figure}
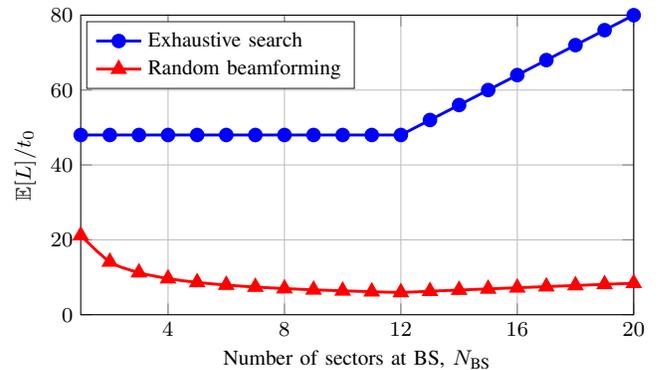

Fig.~\ref{fig:delaycomp} compares the performance of exhaustive search and random beamforming for cell-search, and it shows the normalized expected search latency corresponding to the same search budget $N_c=\Nb \Nu$. As predicted by Remark~\ref{rmk: exhastive-search} and its related discussions, there is a huge gap between the latency of random beamforming and that of exhaustive search.

\begin{figure}
\footnotesize{\begin{tikzpicture}
\begin{axis}[%
width=0.83\columnwidth,
height=0.5\columnwidth,
at={(0\columnwidth,0\columnwidth)},
scale only axis,
xmin=1,
xmax=120,
xlabel={Cell-search latency constraint, $N_c$},
ymode=log,
ymin=1e-07,
ymax=1.01,
ylabel={Detection failure probability, $P_f$},
ylabel near ticks,
xmajorgrids,
ymajorgrids,
yminorgrids,
axis background/.style={fill=white},
legend style={at={(0.01,0.02)},anchor=south west,legend cell align=left, align=left}
]
\addplot [color=red, line width=1.1pt, mark size=2.6pt, mark=triangle*, mark options={solid, fill=red, red}]
  table[row sep=crcr]{%
1	0.9815\\
10	0.82966551560987\\
20	0.688344867792191\\
30	0.571095999654216\\
40	0.47381865701585\\
50	0.393111000378631\\
60	0.326150640821049\\
70	0.270595939583285\\
80	0.224504119736303\\
90	0.207879576350762\\
100	0.207879576350762\\
110	0.207879576350762\\
120	0.207879576350762\\
};
\addlegendentry{RB, $\lambda = 10^{-4}$}

\addplot [color=red, line width=1.1pt, mark size=2.1pt, mark=square*, mark options={solid, fill=red, red}]
  table[row sep=crcr]{%
1	0.8324\\
11	0.132939116122031\\
20	0.0255059457009536\\
29	0.0048936181093819\\
38	0.000938898658424382\\
48	0.000149947570614005\\
57	2.87692193662548e-05\\
67	4.59461147773993e-06\\
76	8.81530690791105e-07\\
86	1.50701727539007e-07\\
100	1.50701727539007e-07\\
110	1.50701727539007e-07\\
120	1.50701727539007e-07\\
};
\addlegendentry{RB, $\lambda = 10^{-3}$}

\addplot [color=blue, line width=1.1pt, mark size=2.2pt, mark=*, mark options={solid, fill=blue, blue}]
  table[row sep=crcr]{%
1	1\\
10	1\\
19	1\\
28	1\\
37	1\\
47	1\\
48	0.000149947570614005\\
57	0.000149947570614005\\
66	0.000149947570614005\\
76	0.000149947570614005\\
85	0.000149947570614005\\
95	0.000149947570614005\\
96	1.50701727539007e-07\\
107	1.50701727539007e-07\\
120	1.50701727539007e-07\\
};
\addlegendentry{EH, $\lambda = 10^{-3}$}
\end{axis}
\end{tikzpicture}%
}

\caption{Tradeoff between the detection failure probability and average latency. ``RB'' and ``EH'' stand for random beamforming and exhaustive search, respectively.}
\label{fig:dfpcomp}
\end{figure}
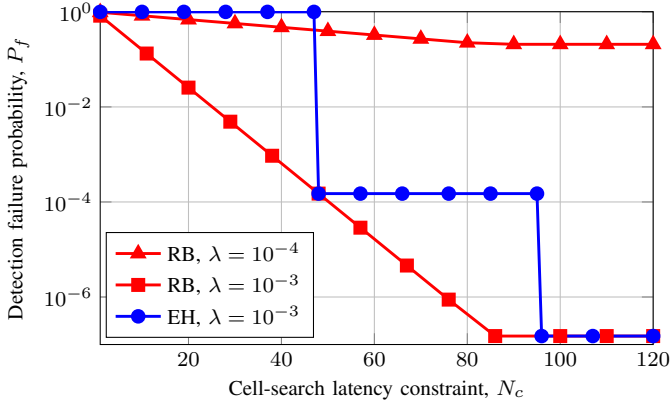

Fig.~\ref{fig:dfpcomp} illustrates the tradeoff between detection failure probability and the cell-search budget. The area above the lines illustrate the feasible performance regions of cell-search, namely there are some settings for $\Nb$ and $\Nu$ that allow realizing any $(\mathbb{E}[L], P_f)$ point above the line. For the exhaustive search, the feasible region is a sequence of step functions due to its quantized latency, which is a subset of that of the random beamforming. Moreover, denser mmWave networks have a larger feasible region, which gives more flexibility to optimize the tradeoff between detection probability and latency.

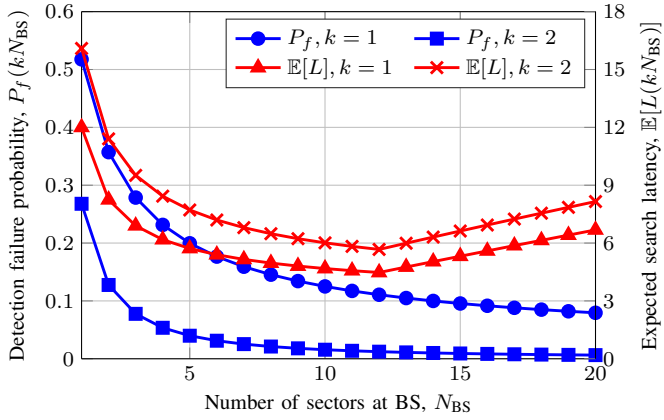
\begin{figure}
\footnotesize{\begin{tikzpicture}
\pgfplotsset{
width=0.95\columnwidth,
height=0.7\columnwidth,
at={(0\columnwidth,0\columnwidth)},
xmin=1,
xmax=20,
xlabel={Number of sectors at BS, $\Nb$},
xmajorgrids,
legend style={at={(0.98,0.98)},anchor=north east,legend cell align=left,align=left,legend columns=2,draw=white!15!black}
}

\begin{axis}[
axis y line*=left,
ymin=0,
ymax=0.6,
ytick={0, 0.1, 0.2, 0.3, 0.4, 0.5, 0.6, 0.7, 0.8},
ylabel={Detection failure probability, $P_f(k\Nb)$},
ylabel style={font=\color{blue}},
ymajorgrids,
ylabel near ticks,
]
\addplot [color=blue, line width=1.1pt, mark size=2.2pt, mark=*, mark options={solid, fill=blue, blue}]
  table[row sep=crcr]{%
1	0.517567371465138\\
2	0.357154082746551\\
3	0.278630046233136\\
4	0.231396701533809\\
5	0.19959214235009\\
6	0.176596010211737\\
7	0.159131271264702\\
8	0.145380964595134\\
9	0.134252486168176\\
10	0.125047413920839\\
11	0.117297469291138\\
12	0.110676399806029\\
13	0.104949621062503\\
14	0.0999439488283727\\
15	0.0955286279834466\\
16	0.0916030136035383\\
17	0.0880883254298338\\
18	0.0849219819474018\\
19	0.0820536163990641\\
20	0.0794422182291809\\
};\label{Plot_one}
\addplot [color=blue, line width=1.1pt, mark size=2.1pt, mark=square*, mark options={solid, fill=blue, blue}]
  table[row sep=crcr]{%
1	0.267875984005332\\
2	0.12755903882253\\
3	0.0776347026638795\\
4	0.0535444334807266\\
5	0.0398370232878984\\
6	0.0311861508227039\\
7	0.0253227614943202\\
8	0.0211356248666117\\
9	0.0180237300423362\\
10	0.0156368557282896\\
11	0.0137586963021055\\
12	0.0122492654740239\\
13	0.011014422961163\\
14	0.00998879290740838\\
15	0.00912571876439974\\
16	0.00839111210125002\\
17	0.0077595530770323\\
18	0.00721174301787484\\
19	0.00673279596416476\\
20	0.0063110660371728\\
};\label{Plot_two}
\end{axis}

\begin{axis}[
axis y line*=right,
axis x line=none,
ymin=0,
ymax=18,
ytick={0,3,6,9,12,15,18},
ylabel={Expected search latency, $\mathbb{E}[L(k\Nb)]$}
]
\addlegendimage{/pgfplots/refstyle=Plot_one}\addlegendentry{$P_f, k=1$}
\addlegendimage{/pgfplots/refstyle=Plot_two}\addlegendentry{$P_f, k=2$}

\addplot [color=red, line width=1.1pt, mark size=2.6pt, mark=triangle*, mark options={solid, fill=red, red}]
  table[row sep=crcr]{%
1	12\\
2	8.24442225491421\\
3	6.89716690802349\\
4	6.17737718607093\\
5	5.71856545418664\\
6	5.39530054952166\\
7	5.15239930851391\\
8	4.96154088947166\\
9	4.80657985388531\\
10	4.67757957047407\\
11	4.56805734102252\\
12	4.47358506144117\\
13	4.75694236121057\\
14	5.03775881622325\\
15	5.3162879415974\\
16	5.59274555243154\\
17	5.86731710960188\\
18	6.14016332249372\\
19	6.41142448863316\\
20	6.68122390495477\\
};
\addlegendentry{$\mathbb{E}[L], k=1~~$}

\addplot [color=red, line width=1.1pt, mark size=3.2pt, mark=x, mark options={solid, fill=red, red}]
  table[row sep=crcr]{%
1	16.0926080610085\\
2	11.4023901286026\\
3	9.5121223164685\\
4	8.43234539814021\\
5	7.71516548485114\\
6	7.19638784214259\\
7	6.79981863237783\\
8	6.48467740829645\\
9	6.22692484202867\\
10	6.01136243907382\\
11	5.82785580130784\\
12	5.6693580137649\\
13	5.99170008030061\\
14	6.30983752995009\\
15	6.62426783655397\\
16	6.93540236025612\\
17	7.24358546696672\\
18	7.54910857430754\\
19	7.85222066356803\\
20	8.15313628502649\\
};
\addlegendentry{$\mathbb{E}[L], k=2$}
\end{axis}

\end{tikzpicture} }

\caption{Detection failure probability and  expected search latency against number of scan cycles $k$ and number of BS sectors $\Nb$.}
\label{fig:dfpdelay}
\end{figure}

Fig.~\ref{fig:dfpdelay} shows the objective and constraint functions of optimization problem~\eqref{eq: optimization-problem}. Increasing $\Nb$ means decreasing $\Thetab$. As shown in the figure, the detection failure probability is a decreasing function of $\Nb$ and therefore narrower beams are always beneficial. The floor of $P_f$ is due to both the blockage and deep fading, which can be improved either by increasing the BS density ($\lambda$) or search budget ($k\Nb$). Meanwhile, as long as the mini-slot duration $t$ is a decreasing function of $\Nb$, namely narrower beams can reduce mini-slot duration, $\mathbb{E}[L(N_c)]$ is a decreasing function of $\Nb$ as well. After a critical point where $t=t_0$, the search latency increases with $\Nb$. In this situation, narrower beams cannot reduce the mini-slot duration anymore but linearly increase $\Nb$ so linearly increases the search space. The optimal beamwidth (or equivalently the optimal $\Nb$) depends on the maximum allowable detection failure probability $P_f^{\text{max}}$.



\section{Conclusion and Future Works}\label{sec: conclusions}
In this paper, we investigated the performance of random beamforming in initial cell-search of mmWave networks. We developed an analytical framework leveraging stochastic geometry to evaluate the detection failure and latency performance. Numerical results showed that random beamforming, though being very efficient from signaling and computational perspectives, can provide near optimal detection and latency performance, especially in dense BS deployment scenarios. Consequently, it may be unnecessary to develop complex algorithms for cell search process in future dense mmWave networks. Meanwhile, random beamforming scheme can be selected as a new benchmark in future initial access studies considering its simplicity and good performance.

The current work focused on LoS links and control plane performance. Future extensions can incorporate NLoS path-loss models and the performance of the data plane.

\appendices
\section*{Appendix\\Proofs}
\subsection{Lemma~\ref{lem: pfasymptotic}}
Define by  $B(o,R)$ a ball of radius $R$ centered at the origin, by $n_R$ the number of BSs in $B(o,R)$, and by $m_R$ the number of BSs in $B(o,R)$ that has LoS to the typical UE. Clearly, $m_R \leq n_R$. Given the Poisson distribution of the BSs, we have
\begin{align}
\Pr(m_R = 0) &= \sum\limits_{n=0}^{\infty} \Pr(m_R = 0\mid n_R = n) \Pr(n_R = n) \nonumber \\
&\stackrel{(a)}{=} \sum\limits_{k=0}^{\infty} \(\int\limits_{0}^{R} (1-e^{-\beta r}) \frac{2r}{R^2} \mathrm{d}r\)^n \frac{e^{-\lambda \pi R^2} (\lambda \pi R^2)^n}{n!} \nonumber \\
& = \exp\(-\frac{2\lambda \pi(1-(\beta R +1)e^{\beta R})}{\beta^2} \) \:,
\end{align}
where (a) follows from the properties of PPP, given a fixed number of points in an area, the points are independently and uniformly distributed over the area, and the the assumption of independent blockages on each link. By setting $R$ to infinity, we extend the ball to the whole plane
\begin{equation*}
P_{\text{no-LoS}} = \lim\limits_{R\to \infty} \Pr(m_R = 0) = \exp\(-\frac{2 \pi\lambda}{\beta^2}\) \:.
\end{equation*}

\subsection{Proposition~\ref{prop: detection-failure-prob}}
\label{sec:P1}
Denote $\text{BS}_{i}$ as the BS located at $x_i$, $r = \lVert x_i \rVert$ and $I_{x_i} \triangleq \sum\limits_{x_j \in \Phi \backslash x_i} h_{j} \ell(\lVert x_j \rVert)  S_{j}$. Then, the successful detection probability in one mini-slot under strongest BS association can be derived as follows:
\begin{align}
\label{equ:Ps}
P_s &= \Pr\(\max\limits_{x_i \in \Phi} \text{SINR}_{x_i} \geq T\) = \Pr\(\bigcup_{x_i \in \Phi} \text{SINR}_{i} \geq T \) \nonumber \\
&\stackrel{(a)}{=} \mathbb{E} \[\sum\limits_{x_i \in \Phi} \mathbbm{1}(\text{SINR}_{i}\geq T)\] \nonumber\\
&\stackrel{(b)}{=} \frac{\Thetab}{2 \pi} \lambda \int_{\mathbb{R}^2} \Pr(\text{SINR}_{i} \geq T \mid r) \mathrm{d}r \nonumber\\
&= \frac{\Thetau\Thetab}{2 \pi} \lambda \int_{0}^{\infty} \hspace{-2.5mm} \Pr(\text{SINR}_{i} \geq T \hspace{-0.8mm} \mid \hspace{-0.8mm} S_i=1, r) \hspace{-0.8mm} \Pr(S_i=1 \mid r) r\mathrm{d}r \nonumber\\
&\stackrel{(c)}{=}  \frac{2 \pi}{\Nb \Nu} \lambda \int_{0}^{\infty} \mathcal{L}_{I_{x_i}} (T r^{\alpha}) e^{-T r^{\alpha} \sigma^2} e^{-\beta r} r\mathrm{d}r \:,
\end{align}
where (a) follows from Lemma 1 in \cite{Dhillon2012} \begin{footnote}{Note that Lemma 1 in \cite{Dhillon2012} is based on $T>1$ (0dB). It also provides a tight upper bound until $T=0.4$ (-4dB).}\end{footnote}, (b) follows from Campbell Mecke Theorem \cite{Chiu2013} and (c) follows the Rayleigh fading assumption. The use of $\Thetau$ and $\Thetab\lambda/2\pi$ are due to BS and UE beamwidth. Here $\mathcal{L}_{I_{x_i}}(T r^{\alpha})$ is the Laplace transform of the interference $I_{x_i}$. Denote $R_j$ as the distance from the $j$th interfering BS to the typical UE, $\mathcal{L}_{I_{x_i}}(T r^{\alpha})$ can be expressed as:

\begin{align}\label{eq:Pc}
\mathcal{L}_{I_{x_i}} & (T r^{\alpha}) = \mathbb{E}_{\Phi, h_i}\[\exp \(-T r^{\alpha} \sum\limits_{x_j \in \Phi \backslash x_i} R_{j}^{-\alpha} h_j S_j \) \]  \nonumber \\
&\stackrel{(a)}{=} \mathbb{E}\[\prod\limits_{x_j \in \Phi \backslash x_i} \hspace{-3mm} \mathbb{E}_{h_j}\[\exp(-T r^{\alpha} R_j^{-\alpha} h_j)\]e^{-\beta R_j} + 1 - e^{\beta R_j} \] \nonumber \\[-0.5mm]
&\stackrel{(b)}{=} \mathbb{E}\[\prod\limits_{x_j \in \Phi \backslash x_i} 1-\frac{T r^{\alpha}e^{-\beta R_j}}{R_j^{\alpha}+T r^{\alpha}}\]  \nonumber \\
&\stackrel{(c)}{=} \exp\(-\Thetau\frac{\Thetab}{2\pi} \lambda_b \int_0^{\infty} \frac{T r^{\alpha}e^{-\beta v}}{v^{\alpha}+T r^{\alpha}} v\mathrm{d}v\) \:,
\end{align}
where (a) follows that $S_j$ is a Bernoulli random variable with parameter $e^{-\beta R_i}$, (b) follows that $h_j$ is an exponential random variable and (c) is derived from the probability generating function of the PPP. Substituting \eqref{eq:Pc} into \eqref{equ:Ps} we obtain the successful detection probability in one mini-slot. Thus the detection failure probability after $N_c$ mini-slots is $(1-P_s)^{N_c}$.

As we have shown in the numerical results, $(1-P_s)^{N_c}$ is very tight for realistic BS density values (e.g., more than 1 BS in every 10000 m$^2$). For a very sparse BS deployment, however, this equation is not valid as it implies $P_f \to 0$ as $N_c \to \infty$ regardless of other parameters. To fix this problem, we note that
\begin{align*}
P_f & = \lim_{R \to \infty}P_{f|m_R = 0} \Pr(m_R = 0) + P_{f|m_R >0} \Pr(m_R >0) \\
& \stackrel{(a)}{\geq}\lim_{R \to \infty} \Pr(m_R = 0) = P_{\text{no-LoS}} \:,
\end{align*}
$P_{f|m_R = m}$ is the failure probability given the existence of $m$ LoS BSs in $B(0,R)$, and (a) follows from that $P_{f|m_R = 0} = 1$ and that $P_{f|m_R >0} \Pr(m_R >0) \geq 0$ for any $R>0$. This completes the proof.

To elaborate on the lower-bound, by setting $N_c \to \infty$, channel fading cannot contribute to the failure probability anymore. In this case, $P_f(\infty)$ is solely due to having no close-enough LoS BS to the typical UE, which is $P_{\text{no-LoS}}$ in Lemma~\ref{lem: pfasymptotic}. In other words, $P_{f|m_R >0} \Pr(m_R >0) \to 0$ as $\lambda \to 0$ and $N_c \to \infty$, making the lower-bound very tight, asymptotically.

\subsection{Proposition~\ref{prop: expected-latency}}
\label{sec:P2}
Denfine event $C$ as $$C = \{\text{A UE can be detected within $N_c$ mini-slots}\},$$ thus $\Pr(C) = 1-P_f(N_c)$. Denote $n_e$ as the number of mini-slots after which the typical UE can be detected. Since each BS chooses a direction uniformly, we have the probability mass function for $n_e>0$ as:
\begin{equation*}
\Pr[n_e = n \mid C] =  (1-P_s)^{n-1} P_s \:.
\end{equation*}
Therefore, the average number of mini-slots for discovering the UE is
\begin{align*}
\sum\limits_{n=1}^{N_c} & n \Pr(n_e = n \mid C) \\
& = \frac{1}{1-P_f} \sum\limits_{n=1}^{N_c} n (1-P_s)^{n-1} P_s \\
& = \frac{1-(N_c+1)(1-P_s)^{N_c}+N_c (1-P_s)^{N_c + 1}}{(1-P_f)P_s} \:.
\end{align*}
This is indeed the normalized expected cell-search latency.

\bibliographystyle{./Components/MetaFiles/IEEEtran}
\bibliography{./Components/MetaFiles/References}

\end{document}